# Defect induced $d^0$ ferromagnetism in a ZnO grain boundary

Assa Aravindh S, Udo Schwingenschloegl and Iman S Roqan
Division of Physical Sciences and Engineering
King Abdullah University of Science and Technology
Thuwal 2955-6900, Kingdom of Saudi Arabia

Several experimental studies have referred to the grain boundary (GB) defect as the origin of the ferromagnetism in ZnO. However, the mechanism of this hypothesis has never been confirmed. Present study investigates the atomic structure and the effect of point defects in a ZnO GB using the GGA+$U$ approximation. The relaxed GB possesses large periodicity and channels with 8 and 10 numbered atoms having 4 and 3 fold coordination. Unlike O vacancy ($V_O$), Zn vacancy ($V_{Zn}$) is preferentially aggregated at the GB, relative to the bulk-like region, indicating the possibility of obtaining p-type conductivity in polycrystalline ZnO. Although no magnetization is obtained from point defect-free GB, $V_{Zn}$ induces spin polarization as large as 0.68 $\mu_B$/atom to the O sites at the GB. Ferromagnetic exchange energy > 150 eV is obtained by increasing the concentration of $V_{Zn}$ and by the injection of holes into the system. Electronic structure analysis indicates that the spin polarization without external dopants originates from the O 2$p$ orbitals, a common feature of $d^0$ semiconductors.

# 1. Introduction

Recent years have witnessed enormous interest in zinc oxide (ZnO), due to its potential optoelectronic and spintronic applications, such as in light emitting diodes and spin-polarized solar cells, as ZnO offers the possibility of combining semiconducting and magnetic effects [1-3]. The realization of these technologies depends on the development of room temperature ferromagnetism (RTFM) in ZnO. Even though doping ZnO with transition metals (TMs) is a widely employed technique, [4] many studies failed to reproduce the long-range RTFM [5, 6]. In some extant studies, the observed FM has been attributed to the secondary phases and clusters formed in ZnO [7]. Irrespective of all these controversies, it has been widely accepted that a subtle balance between many types of defects plays a key role in the RTFM of ZnO. Several authors have reported spin polarization in ZnO induced by cation vacancies [8, 9], which arises from localized holes forming triplet states at a stable magnetic ground state. Since these localized holes are located in the $p$-band, rather than in the $d$-band, this phenomenon is referred to as $d^0$ FM [10-13]. It has been proven to be an alternative to TM doping, as the magnetization originates in the presence of sufficient holes. However, the formation energy ($E^f$) of cation vacancy in bulk ZnO is high, making the achievement of FM difficult. One of the methods suggested to reduce the $E^f$ of cation vacancies in bulk ZnO is a co-doping approach[13]. In another study, Singh et al[12] have found that cation vacancies can be easily formed in ZnO nano clusters whereby $d^0$ magnetism originate from O $2p$ orbitals due to the dangling bonds created by these cation vacancies. Nevertheless, intrinsic point defects can be easily formed in ZnO samples, along with other structural defects (dislocations, stacking faults, and grain boundaries (GBs)) and presence of these structural defects is found to be crucial in modifying the electrical, magnetic and optical properties of ZnO [14-16]. Particularly, the coordination number of atoms in a GB is different from that in the bulk and over or under-coordination of atoms can exist at the GBs, which in turn will modulate the properties of polycrystalline materials. The presence of dangling or distorted bonds at the GB can facilitate native defect formations and dopant segregations. The role of GB in promoting RTFM in ZnO is suggested in experimental studies[14, 17, 18]. Magnetic force microscopy has predicted that intrinsic and extrinsic defects at GBs can be the origin of the FM observed in Mn-doped ZnO thin films [17]. However, the contribution of GBs in ZnO in promoting RTFM is not studied in detail and can be understood only through a systematic and microscopic study that incorporates varied concentrations of point defects at the GBs.

The properties of polycrystalline materials are intimately related to the atomic structure of GBs; hence the effect of different GBs on the material properties will be different. Simulations on well-defined GBs are thus essential as it is possible to have wide variety of GB atomic structures, depending on the orientation relationships of the GB planes and the translation states of the two adjacent grains influencing the material behavior [19, 20]. Experimentally it is possible to obtain GBs with low ($\sim <15^0$) [19] and high angles[20, 21] that possess different characteristics. Therefore, the atomic structure of each GB will be unique and the GB properties can be improved by tuning the atomic structure. In this scenario, magnetism in polycrystalline ZnO containing point defects at the GB presents an interesting research problem from the fundamental and technological point of view. This motivated us to investigate the contribution of a GB containing point defects towards the origin of RTFM in ZnO. Zn and O vacancies (henceforth referred to as $V_{Zn}$ and $V_O$, respectively) are introduced at the GB to understand the contribution of GB-defect complexes to the RTFM of otherwise undoped ZnO.

## II. Computational Methodology

The GB is formed by attaching two wurtzite (0001) ZnO crystals oriented at an angle of 45º relative to each other, individually having 148 atoms, such that the supercell contains 296 atoms. The calculations are performed using the pseudopotential method based on density functional theory as implemented in the Vienna Ab-initio Simulation Package within the generalized gradient approximation (GGA) [22, 23]. The pseudopotentials are based on the projector augmented wave formalism [24]. The atomic coordinates are relaxed with energy and force tolerances of 0.001 eV and 0.001 eV/Å, respectively and relaxation is conducted on both neutral and charged supercells. A cut-off energy of 400 eV is used to expand the plane waves included in the basis set and Brillouin-zone sampling is carried out in the Monkhorst-Pack scheme using a 1×1×6 k-point mesh. Owing to the strong electronic correlations of the *3d* electrons, standard density functional theory fails to accurately describe the electronic structure of ZnO. Therefore, the GGA+$U$ method with on-site Coulomb parameter $U = 7$ eV with a $J$ value of 1 eV ($U$-$J = 6$) is used in the Duradev's approach [25]. The choice of $U$ and $J$ is validated by comparing the calculated values of the lattice parameter with experimental values. The calculated bulk unit cell parameters are $a_0 = b_0 = 3.32$ Å and $c_0 = 5.219$ Å, which is in good agreement with experimental findings ($a_0 = b_0 = 3.2507$ Å; $c_0 = 5.2083$ Å [26]). These optimized values of bulk lattice parameters are used for constructing the GB supercell, such that $a=7 \times a_0$, $b=\sqrt{3} \times 3 \times b_0$; and $c=c_0$. Such a large supercell ensures negligible interaction between the periodically repeated GB images. The formation energy ($E^f$) of native defects, such as $V_{Zn}$ and $V_O$, is calculated using the expression [27]:

$$E^f(D) = (E_{tot}(D) - E_{tot}(ZnO) + n_i\mu_i + qE_F)/n_i, \quad (1)$$

Where $E_{tot}(D^q)$ is the total energy of the ZnO supercell containing the defective grain boundary, and $E_{tot}(ZnO)$ is the total energy of same supercell without point defects at the GB. Here, $n_i$ and $\mu_i$ are, respectively, the number and chemical potential (which is dependent on the experimental growth conditions) of the atoms removed from the supercell. Extreme O-rich conditions are represented by the energy of O in the $O_2$ molecule, while that of Zn is given by the energy of bulk Zn. Finally, $q$ denotes the charge state and $E_F$ is Fermi energy. The inclusion of $E_F$ accounts for the various charge states in which defects can occur.

111. Results and Discussion

1) Structure and energetics

The atomic coordinates of the GB are allowed to relax without any constraints and the geometry before and after relaxation is shown in Fig. 1. It can be seen from Fig.1 that the two ZnO single crystals are directly bonded at the GB and no inter-granular structures are formed at the boundary. Structural units in which the Zn and O atoms are 4 fold and 3 fold coordinated appears at the GB after relaxation, in contrast to the bulk ZnO having 4 fold coordination. The relaxed GB is stable with no Zn-Zn or O-O "wrong bonds" and consists of channels with 8 and 10 membered atoms, compared to the 6 atom open channel in bulk ZnO. The 8 and 10 atom ring patterns are repeated, indicating that the atomic arrangement is periodic at the GB. A closer look at the bond lengths at the GB indicates that, after optimization, the Zn-O bonds undergo slight change of ~ 0.05 Å in the 8 and 10 membered rings compared to the bulk, to minimize the local forces in the vicinity of the GB.

To investigate the effect of the presence of intrinsic point defects at the GB on the structural, electronic and magnetic properties, we introduce single $V_{Zn}$ and $V_O$ at the GB, with atomic relaxation carried out for each case. This amounts to a vacancy concentration of 0.34% in the ZnO supercell. The $E^f$ of these defects is calculated according to Equation (1). Although it is known that $V_O$ is more stable than $V_{Zn}$ in bulk ZnO [27], $E^f$ shows that the opposite is true at the GB. Since the $45^0$ GB has relatively long structural periodicity along the GB plane, a systematic investigation on the preference of atomic site of vacancies is inevitable. Motivated by the low $E^f$ and favorable magnetism exhibited by $V_{Zn}$ we have analyzed 14 Zn vacancy sites at the GB as shown in Fig. 2. It can be seen that the most favorable $V_{Zn}$ position is located at a neck site in the 8 atom membered ring ($V_7$) with 3-fold coordination (The $E^f$ values for all $V_{Zn}$ positions ($V_1$ to $V_{14}$) are indicated in Fig. 2).

The Zn-O bond length contraction near $V_7$ is ~ 0.1 Å, in average, to adjust the local strain in the vicinity of the GB. We find that $V_{11}$ (3.8 eV) and $V_{15}$ (4.07 eV) also possess very low $E^f$ with 3-fold coordination at the GB. Hence, it is easier to break these bonds relative to those formed at 4 fold-coordinated sites in the GB. For $V_O$, five different positions are considered and their respective $E^f$ values are given in Fig. 3. For comparison, $E^f$ is also calculated for $V_{Zn}$ and $V_O$ located in the bulk-like region away from the GB. Interestingly, $V_O$ also forms more easily in the GB ($E^f$ = 4.1eV; see Fig. 3), compared to bulk-like region ($E^f$= 5.3eV). However, $V_{Zn}$ is also more stable at the GB with $E^f$ = 3.69 eV (see Fig. 2) relative to the bulk-like region ($E^f$ = 5.10 eV). We find that $E^f$ of $V_{Zn}$ GB is lower than in bulk ZnO ($E^f$ = 5.4eV) [28, 29]. This finding indicates that the GB can stabilize $V_{Zn}$. Creating $V_{Zn}$ by equilibrium or non-equilibrium processes yields *p*-type conductivity in ZnO. Therefore, this is in accordance with previous observations where *p*-type conductivity in ZnO has been attributed to the impurities and vacancies at the GBs [30]. Even though the $E^f$ of these intrinsic vacancies is high, they are unavoidable during growth, depending on the growth conditions and thermal fluctuations that occur during the process. In electron paramagnetic resonance (EPR) [31] and positron annihilation spectroscopy (PAS) [32] experiments, $V_{Zn}$ has been detected with large thermal stability in ZnO samples. Our results indicate that engineering GBs in polycrystalline samples assists in stabilizing $V_{Zn}$ in ZnO.

A pair of $V_{Zn}$ is introduced at the GB to study the effect of vacancy concentration on the structural and magnetic properties, such that the vacancy concentration now amounts to 0.68% in the ZnO supercell. Different pair configurations are considered to examine the vacancy stability and magnetic coupling. To create a pair, we combine $V_7$ (the most stable site) with a second vacancy (positions shown in Fig. 2), which is chosen based on the distance between the two vacancies and energetic preference. We study the effect of the distance of separation between the vacancy pairs ($D$) to understand the aggregation tendency. The $E^f$ of different $V_{Zn}$ pairs in the GB with their respective $D$ values is presented in Table 1. The vacancy pairs are generally more stable at the 3 fold coordinated sites, compared to those with 4-fold coordination. In particular, the vacancy pairs $V_7$-$V_{11}$ and $V_7$-$V_{15}$ that are located at the neck sites with 3-fold coordination are the most stable ones. In addition, the second vacancy prefers to be a first neighbor, as bond relaxations have already occurred in the vicinity of the first vacancy. It is seen that the $E^f$ increases with increase in $D$, indicating that the special bond arrangements at the GB promote vacancy aggregation. This vacancy clustering may affect the electrical, optical and magnetic properties of the material. These results suggest that polycrystalline ZnO can be a reproducible p-type material compared to single crystal ZnO.

2) Magnetism

No spin polarization is observed from either the point defect-free GBs or from introduction of a single $V_O$ at both the GB and at the bulk-like region. Furthermore, we introduce a $V_O$ pair to explore the possibility of producing magnetism via an increased anion vacancy concentration, obtaining zero spin polarization. On the other hand, introducing $V_{Zn}$ at the GB induces magnetic moments on the nearest neighbor O atoms, whereas the magnetic moments on Zn atoms are negligible. To visualize the magnetism induced by $V_{Zn}$, we plot the spin density distribution at the GB for the stable $V_{Zn}$ (Fig. 4). Difference between the majority and minority spin densities is calculated and indicated by iso-surfaces near $V_{Zn}$ and at the vacancy center. The iso-surfaces show extended tails in the immediate vicinity of the vacancy. It can be seen from Fig. 4 that the effect of spin polarization induced by $V_{Zn}$ is not limited to the nearest neighbor O atoms and extends to several O atoms near and at the GB. In addition, we find that the under-coordinated O atoms created by the neutral $V_{Zn}$ are easily polarized and the magnetic moments are not symmetrically distributed among O atoms due to the different spatial arrangement of atoms at the GB. The local moments of the O atoms in the vicinity of $V_{Zn}$ range from 0.3 to 0.68$\mu_B$/atom. These O atoms with unsatisfied valence have dangling bonds and give rise to residual spin polarization due to the unpaired electrons. Hence, the O atoms possessing dangling bonds are magnetized more easily compared to other O atoms. Such magnetism is the characteristic of a $d^0$ semiconductor, as it is induced by the cation vacancy due to the spatial local character of the *2p* states of the nearest anion sites around/near the defect center [10, 33, 34]. In contrast, the magnetic moment of the O atoms surrounding the $V_{Zn}$ in the bulk-like region is smaller and limited to the nearest neighbor atoms. This emphasizes the role of cation vacancies at the GBs in promoting the FM observed in TM or RE-doped ZnO [11-15].

To further investigate the possibility of RT ferromagnetic coupling originating from the GB, we calculate the energy difference ($\Delta E$) between the ferromagnetic and antiferromagnetic (AFM) states for the two vacancy configurations and shown in Table 2. The exchange coupling energy difference between the two vacancies ($V_{Zn1}$ and $V_{Zn2}$) is defined as:

$$\Delta E (V_{Zn1} - V_{Zn2}) = E (V_{Zn1\uparrow}\ V_{Zn2\downarrow}) - E (V_{Zn1\uparrow} V_{Zn2\uparrow}) \quad (2)$$

where the arrows represent spin-up and spin-down states, respectively. A positive value of $\Delta E$ indicates that FM is favored over AFM and vice versa. $\Delta E$ is calculated for neutral as well as for +1 and +2 charge states, and presented in Table 2. Holes are introduced by reducing the total number of electrons in the system. The removal of one and two electrons represents the injection of one and two holes into the system respectively[22, 23]. We have also tested electron injection by increasing the total number of electrons, however no enhancement in the FM is observed. It is well established that defect-induced FM requires the band edge hybridization with the defect bands near the $E_F$. This, in turn, facilitates the magnetic exchange coupling between the host carriers and defect states. Hole injection enhances the hybridization of states near VB by moving the $E_F$ towards the VBM; thus enhancing the magnetic coupling[35]. On the other hand, increasing electron carrier density is expected to move the $E_F$ away from the VBM, into the conduction band[36], reducing the hybridization near the $E_F$ whereby no ferromagnetic enhancement can take place. We find that weak ferromagnetic coupling is favored for the neutral charge state for both configurations, with small energy differences (ΔE < 30 meV). However, the introduction of a positive charge enhances the ΔE significantly, favoring strong ferromagnetic coupling between the vacancies. We observe that ΔE increases with double positive charge (155 and 187 meV for $V_7$-$V_{11}$ and $V_7$-$V_{15}$, respectively) compared to single charge. This energy difference is sufficient to stabilize FM against thermal fluctuations, since ΔE > 30 meV is the minimum required to establish ferromagnetic coupling at RT [37] (especially as the $E^f$ shows that presence of holes increases the stability of vacancy pairs). Interestingly, the total magnetic moment of the GB supercell also increases with the introduction of holes into the system. Although no change in the magnitude of the maximum local magnetic moment is observed with the increase in $V_{Zn}$ concentration, the net magnetic moment of the system increases from ~ 3 to 4.5 $μ_B$ due to the presence of increased hole carriers, as can be seen from Table 2. The increase in the concentration of $V_{Zn}$ and holes at the GB raises the local hole concentration at the anion sites, shifting the $E_F$ towards the valence band maximum (VBM). As a result of this collective phenomenon, hybridization of O $2p$ and Zn $3d$ states near $E_F$ increases, enhancing the ferromagnetic exchange interactions. This finding shows that RTFM can be obtained in ZnO with cation vacancies located at the GB, while simultaneously maintaining sufficient hole concentration [33].

3) Electronic structure

Density of states (DOS) analysis is carried out to gain further insight in to the observed effect of the vacancies at the GB. Initially, we calculate the total density of states (TDOS) of bulk ZnO using GGA+$U$ method (see the top panel of Fig. 5). The TDOS shows the semiconducting and nonmagnetic nature of bulk ZnO and the obtained band gap of 2.5 eV is a significant improvement compared to the values obtained using GGA [28]. The TDOS for the GB supercell without any vacancies is also calculated using the GGA+$U$ scheme, as shown in Fig. 5 (bottom panel). The spin-up and spin-down DOS are equal with $E_F$ positioned in the band gap, demonstrating a semiconducting and non-magnetic nature of the GB. The DOS peaks pertaining to the GB occur at different positions compared to the bulk, due to the difference in bonding characteristics. We find that the band gap calculated from the TDOS of ZnO with GB is smaller than that of bulk ZnO. The change in coordination number of the atoms (8 and 10 atom membered rings with 3 and 4 fold coordination) at the GB compared to the bulk (6 atom membered ring with 4 fold coordination) affects the band structure of the material, owing to the reduction in covalent nature and hybridization of atoms [38]. The band broadening due to the irregular atomic arrangements at the GB causes this narrowing of the band gap. Therefore, the reduced coordination number, as well as the inhomogeneous changes in bond lengths, causes the difference in band structure, compared to grain-free ZnO. Deep unoccupied states in the band gap are not found for the GB from the DOS calculations [39, 40].

The TDOS of the GB containing single $V_O$ and $V_{Zn}$ (with the lowest $E^f$) is also calculated and presented in Figs. 6(a) and 6(b), respectively. The TDOS of $V_O$ exhibits semiconducting properties, with symmetrical majority (spin-up) and minority (spin-down) bands, resulting in zero magnetization of the system. In contrast, the TDOS of the GB with $V_{Zn}$ shows that the majority spins valence states are fully occupied, while the minority ones are partially occupied. Therefore, an increase in the net spin moment is observed due to the difference between the majority and minority spin densities.

Further, the projected DOS (PDOS) of O *2s*, *2p* and Zn *4s*, *4p*, *3d* orbitals ($t_{2g}$ and $e_g$) are shown in Fig. 7(a-c) for the GB containing $V_{Zn}$. The VBM mainly consists of O *2p* states (Fig. 7(a)). A mixture of *s* and *p* states of Zn and O forms the conduction band minimum (CBM), whereas the contributions of the Zn *4s* and *4p* states are negligible, as can be seen from Fig. 7(b). It is also evident that Zn *3d* states ($t_{2g}$ and $e_g$) contribute to the VBM (Fig. 7(c)). A further analysis of the PDOS indicates that the observed magnetism arises from the O *2p* orbitals surrounding the $V_{Zn}$; however, the contribution of the O *2s* states are negligible. The formation of two holes in the *2p* band at the O sites surrounding the $V_{Zn}$ is the main source of this magnetization, whereas the contributions from the Zn *4s* and *4p* states are negligible. Consequently, the observed magnetism is mainly due to the O *p* states at the top of the VB. However, there is a small contribution from the Zn *3d* states, which can hybridize with the O *p* states, promoting the *p-d* exchange interaction [41]. The origin of magnetism from this *p-d* interaction can be traced back to the oxygen atomic properties. Substitutionally doped O in solids has been shown to possess finite magnetic moments [42]. O atom contains two unpaired *2p* electrons, leading to a net magnetic moment of $2\mu_B$. On the other hand, the maximum local moment at the GB, possessed by the O atoms in the vicinity of $V_{Zn}$, is about 0.68 $\mu_B$/atom. The removal of a Zn atom from the GB results in unpaired dangling bonds and the neighboring O atoms move outward from their original positions, causing a prominent Jahn-Teller type distortion. This reduces the local magnetic moment of the O atoms from the ideal value of $2\mu_B$. However, the total spin polarization of the system increases in the presence of higher density of vacancies and/or other defects located near the GB, which enhance the ferromagnetic exchange interactions and thereby promote ferromagnetic coupling. With the introduction of more number of holes, $E_F$ shifts into the VBM; hence, by Stoner's criterion, the spin polarization increases [43]. On the contrary, if the hole density decreases, $E_F$ remains inside the band gap, resulting in zero net magnetization, as obtained for the $V_O$.

IV. Conclusions

GBs are unavoidable structural elements in ZnO samples, irrespective of the method of preparation. They are often complex defects and, as such, can exhibit characteristics of both intrinsic and extrinsic defects. Our theoretical study demonstrates that, no RTFM can be obtained from the GB without point defects. In a ZnO GB, $V_{Zn}$ forms more easily than $V_O$ and the presence of $V_{Zn}$ at the GB is responsible for the experimentally observed RTFM in polycrystalline ZnO. The FM is further enhanced by the introduction of additional holes into the system. Our study explains that the presence of point defects at the GBs play a dominant role in the FM observed experimentally in ZnO samples.

Table 1. Zn vacancy pairs, separation between them ($D$), $E^f$ and maximum local magnetic moment obtained for the O atom in the vicinity of $V_{Zn}$.

| Vacancy Pair | $D$ (A$^0$) | $E^f$ (eV) | Maximum local magnetic moment of O ($\mu_B$/atom) |
|---|---|---|---|
| $V_7$-$V_1$ | 2.38 | 4.35 | 0.58 |
| $V_7$-$V_5$ | 3.13 | 4.86 | 0.49 |
| $V_7$-$V_3$ | 3.47 | 4.05 | 0.56 |
| $V_7$-$V_{15}$ | 5.28 | 3.87 | 0.60 |
| $V_7$-$V_{11}$ | 6.01 | 3.90 | 0.68 |
| $V_7$-$V_{13}$ | 8.15 | 5.12 | 0.53 |
| $V_4$-$V_{12}$ | 11.23 | 5.55 | 0.58 |

Table 2. Separation between two vacancy pairs ($D$), their charge state, total magnetic moment of the GB supercell and AFM-FM energy difference ($\Delta E$).

| Vacancy Pair | $D$ (Å) | Charge State | Total Mag ($\mu_B$) | $\Delta E$ (meV) |
|---|---|---|---|---|
| V$_7$-V$_{11}$ | 6.01 | 0 | 3.00 | 10 |
|  |  | +1 | 3.53 | 47 |
|  |  | +2 | 4.5 | 155 |
| V$_7$-V$_{15}$ | 5.28 | 0 | 3.2 | 23 |
|  |  | +1 | 4.02 | 56 |
|  |  | +2 | 4.53 | 187 |

Figure Captions

Figure 1. ZnO GB structure before (a) and after (b) relaxation. The grey and pink atoms indicate Zn and O respectively.

Figure 2. Relaxed GB; the examined Zn vacancy positions are indicated with V$_i$ symbols (V$_1$, …V$_{14}$) along with their $E^f$ (in eV). The grey and pink atoms indicate Zn and O, respectively. Note that the most favorable vacancy position is V$_7$ at the GB.

Figure 3. Relaxed GB; the examined O vacancy positions are indicated with symbols V$_1$-V$_5$. The $E^f$ (in eV) is also shown. The grey and pink atoms indicate Zn and O, respectively. Note that the most favorable vacancy position is V$_2$ at the GB.

Figure 4. Spin density of GB with the most stable V$_{Zn}$. The grey and pink atoms indicate Zn and O, respectively.

Figure 5. Total DOS of bulk ZnO and GB supercell. The line at zero indicates $E_F$.

Figure 6. Total DOS of GB with V$_{Zn}$ and V$_O$. The line at zero indicates $E_F$.

Figure 7. Orbital resolved DOS of Zn and O atoms when V$_{Zn}$ is present at the GB. The line at zero indicates $E_F$.

Figure 1.

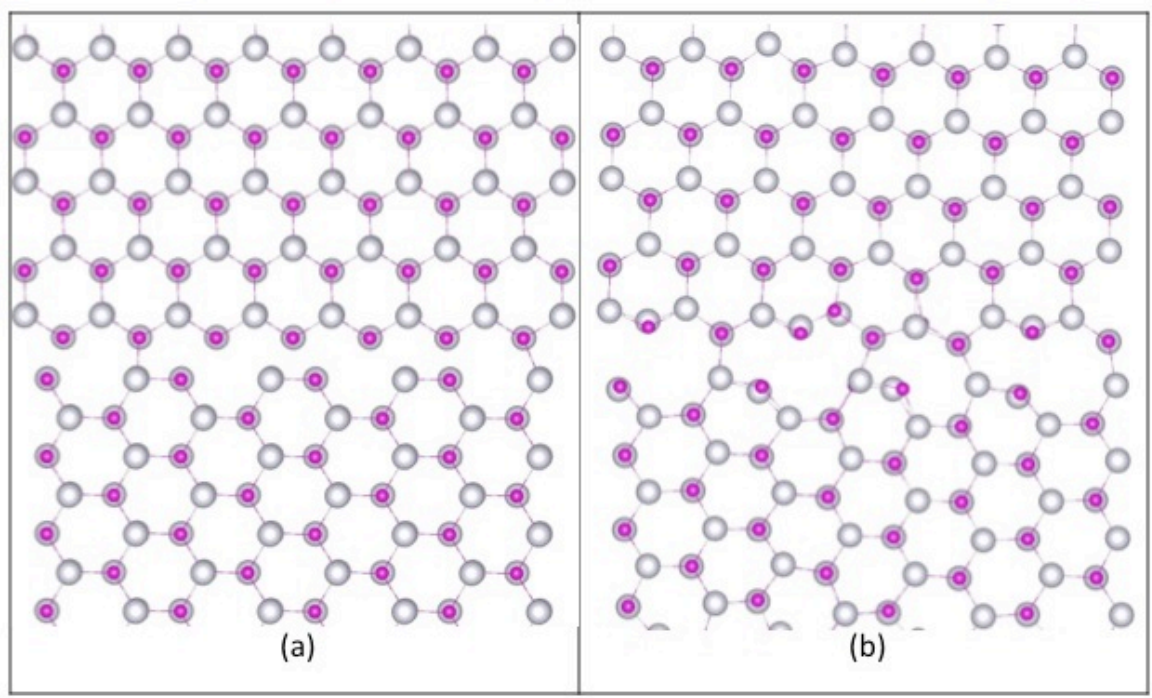

Figure 2.

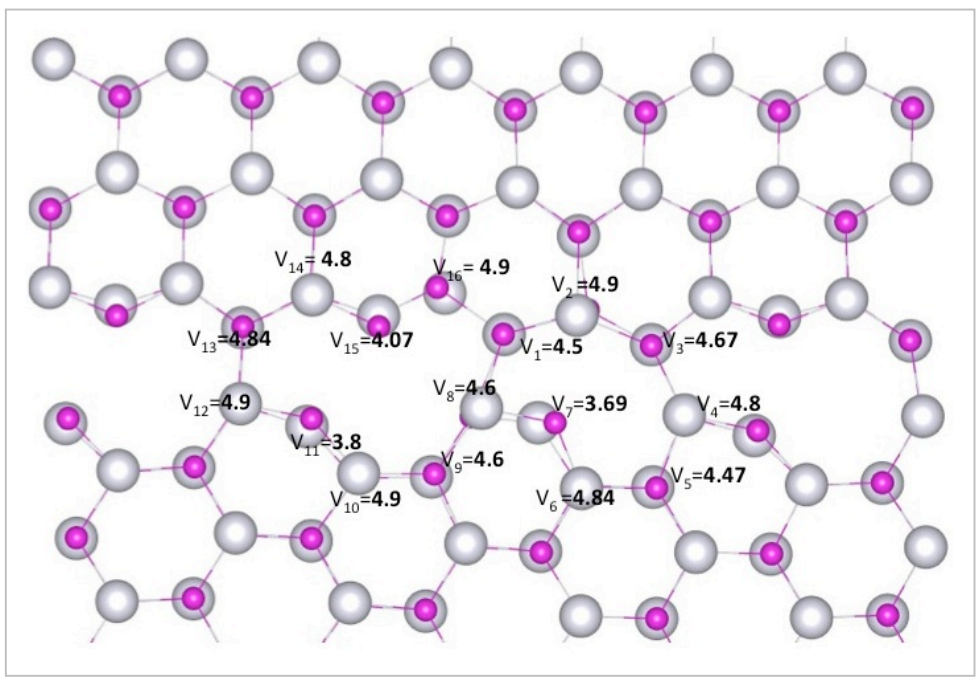

Figure 3.

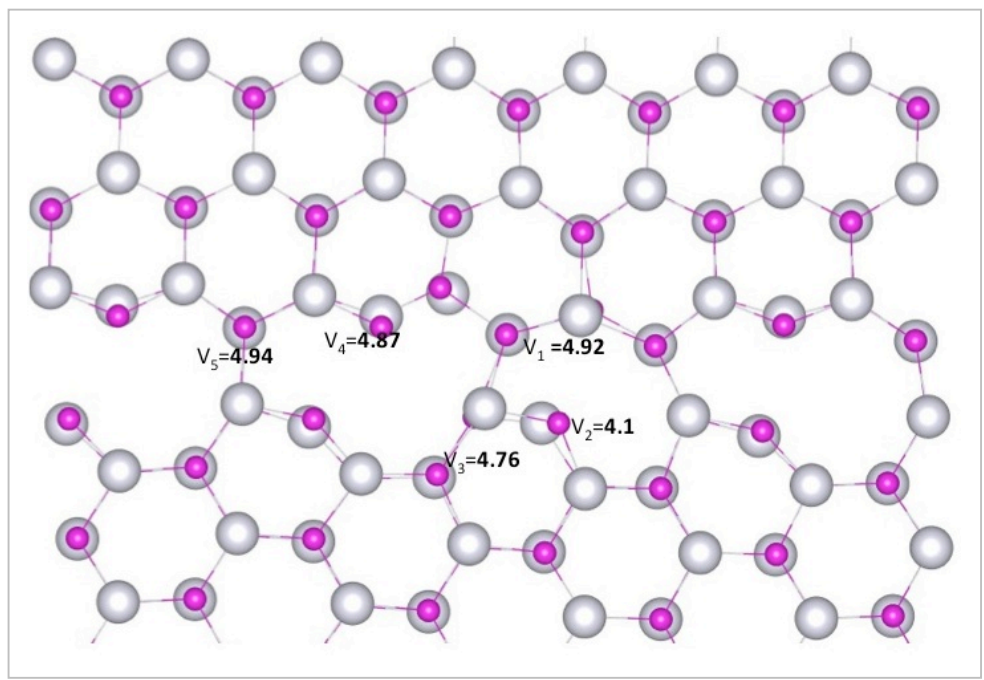

Figure 4.

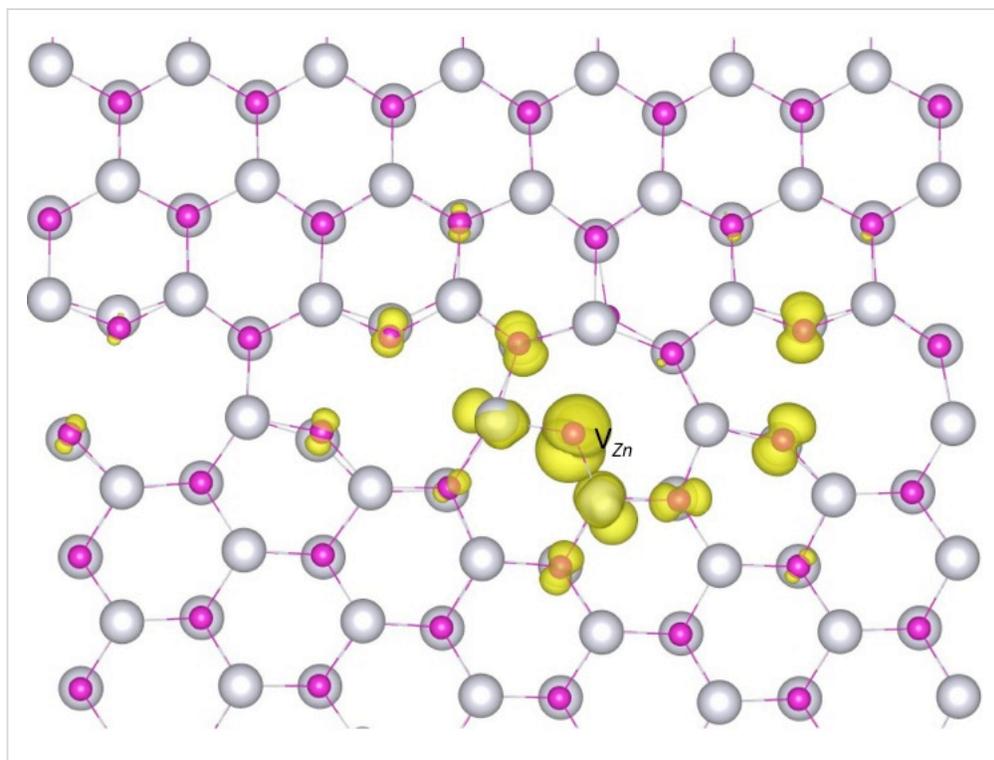

Figure 5.

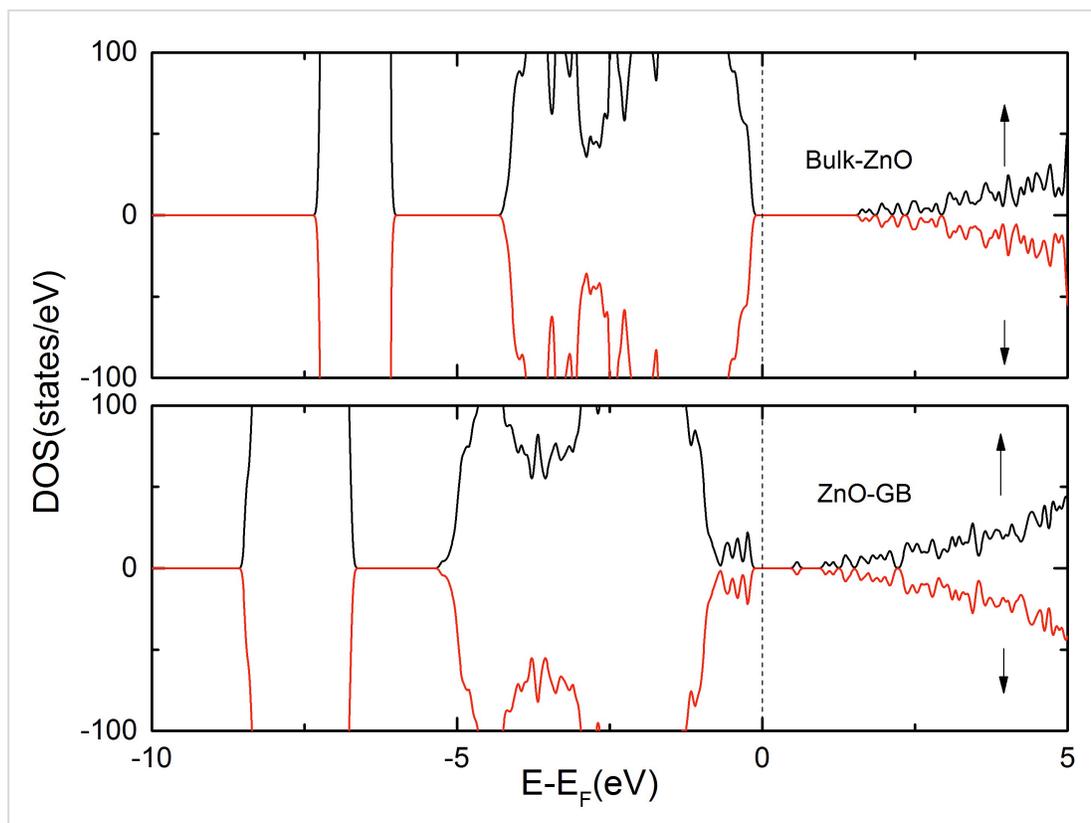

Figure 6.

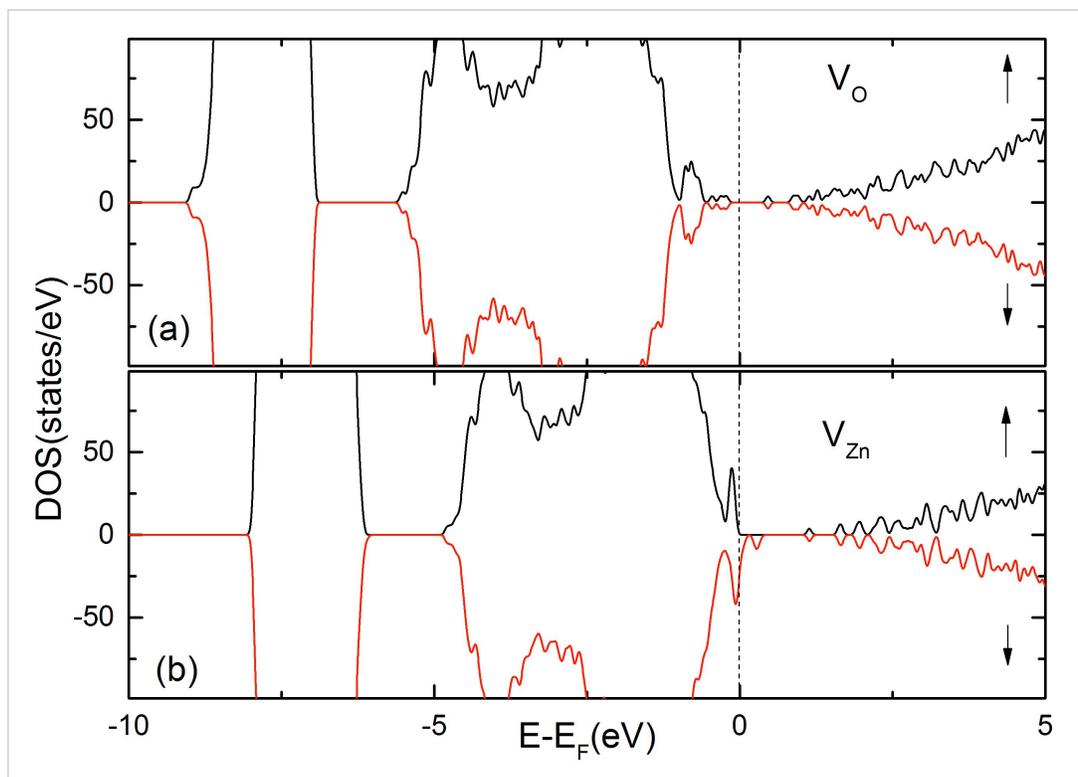

Figure 7.

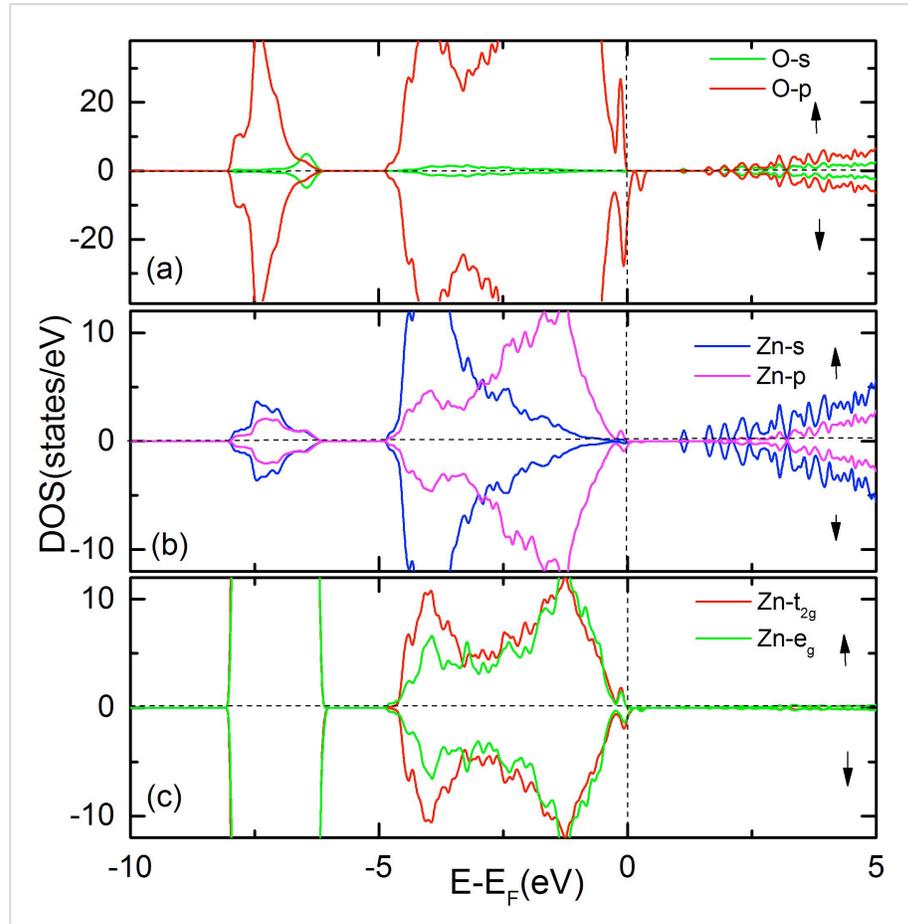